\documentclass[10pt,conference]{IEEEtran}

\usepackage{amsmath}    %
\usepackage{amsfonts}   %
\usepackage{amssymb}    %
\usepackage{algorithm}
\usepackage{algorithmic}
\usepackage{array}
\usepackage{textcomp}
\usepackage{stfloats}
\usepackage{url}
\usepackage{verbatim}
\usepackage{graphicx}
\usepackage{subcaption}
\usepackage{ifthen}
\usepackage{makecell}
\usepackage{xcolor}
\usepackage{float}
\usepackage[most]{tcolorbox}
\usepackage{listings}
\usepackage{tipa}
\usepackage{soul}
\usepackage{fontawesome5}
\usepackage[utf8]{inputenc}
\usepackage{microtype}
\usepackage{balance}
\usepackage[
  colorlinks=true,      %
  linkcolor=black,      %
  citecolor=black,       %
  urlcolor=black         %
]{hyperref}
\definecolor{codegreen}{rgb}{0,0.6,0}
\definecolor{codegray}{rgb}{0.5,0.5,0.5}
\definecolor{codepurple}{rgb}{0.58,0,0.82}
\definecolor{backcolour}{rgb}{0.95,0.95,0.92}
\definecolor{tableblue1}{HTML}{6A93D4}
\definecolor{tableblue2}{HTML}{052D6E}

\lstdefinestyle{mystyle}{
  language=Python,
  backgroundcolor=\color{backcolour},
  basicstyle=\small\ttfamily,
  numbers=left,
  numberstyle=\tiny\color{codegray},
  stepnumber=1,
  numbersep=10pt,
  tabsize=4,
  showspaces=false,
  showstringspaces=false,
  showtabs=false,
  frame=single,
  rulecolor=\color{black},
  aboveskip=0pt,
  belowskip=0pt,
  captionpos=b,
  stringstyle=\color{codepurple},
  keywordstyle=\color{blue},
  commentstyle=\color{codegreen},
  escapeinside={\%*}{*)},
  morekeywords={*,...},
}
\newtcolorbox{boxK}{
    sharpish corners,
    boxrule = 0pt,
    toprule = 2pt,
    enhanced,
    fuzzy shadow = {0pt}{-2pt}{-0.5pt}{0.5pt}{black!35},
    left=0pt,
    right=0pt,
    top=0pt,
    bottom=0pt
}

\usepackage{lstlinebgrd} %

\definecolor{Added}{HTML}{E6F4EA}   %
\definecolor{Removed}{HTML}{FDE7E9} %

\lstdefinestyle{javacode}{
  language=Java,
  basicstyle=\ttfamily\scriptsize,
  numbers=left,
  numbersep=6pt,     %
  xleftmargin=3pt,   %
  numberstyle=\scriptsize\color{gray},
  columns=fullflexible,
  keepspaces=true,
  frame=single,                  %
  backgroundcolor=\color{white},%
  escapeinside={(*@}{@*)}, %
}

\newcommand{\RNum}[1]{\uppercase\expandafter{\romannumeral #1\relax}}
\title{Backdoors in Code Summarizers: How Bad Is It?}

\author{%
  \\[-2em]
  Chenyu Wang\IEEEauthorrefmark{5},
  Zhou Yang\IEEEauthorrefmark{2}\IEEEauthorrefmark{4}\IEEEauthorrefmark{1}\thanks{* Corresponding author.},
  Yaniv Harel\IEEEauthorrefmark{3},
  David Lo\IEEEauthorrefmark{5}\\
  \IEEEauthorblockA{\small \IEEEauthorrefmark{5}Singapore Management University \quad
  \IEEEauthorrefmark{2}University of Alberta \quad
  \IEEEauthorrefmark{3}Tel Aviv University \quad
  \IEEEauthorrefmark{4}Alberta Machine Intelligence Institute}
  \IEEEauthorblockA{\small Email: \{chenyuwang, davidlo\}@smu.edu.sg \quad zhou.yang@ualberta.ca \quad yaniv10@tauex.tau.ac.il}
  \\[-2em]
}

\begin{document}
\pagestyle{plain}

\IEEEoverridecommandlockouts
\maketitle
\thispagestyle{plain}
\begin{abstract}
Large Language Models for Code (Code LLMs) are increasingly employed in software development. 
However, studies have recently shown that these models are vulnerable to \textit{backdoor attacks}: when a \textit{trigger} (a specific input pattern) appears in the input, the backdoor will be activated and cause the model to generate malicious outputs desired by the attacker.
Researchers have designed various triggers and demonstrated the feasibility of implanting backdoors by poisoning a fraction of the training data (known as \textit{data poisoning}).
Some basic conclusions have been made, such as backdoors becoming easier to implant when attackers modify more training data.
However, existing research has not explored other factors influencing backdoor attacks on Code LLMs, such as training batch size, epoch number, and the broader design space for triggers, e.g., trigger length.

To bridge this gap, we use the code summarization task as an example to perform a comprehensive empirical study that systematically investigates the factors affecting backdoor effectiveness and understands the extent of the threat posed by backdoor attacks on Code LLMs.
Three categories of factors are considered: data, model, and inference, revealing findings overlooked in previous studies for practitioners to mitigate backdoor threats.
For example, Code LLM developers can adopt higher batch sizes with fewer epochs appropriately.
Users of code models can adjust inference parameters, such as using a higher temperature or a larger top-k, appropriately.
Future backdoor defense can prioritize the inspection of rarer and longer tokens, since they are more effective if they are indeed triggers.
Since these non-backdoor design factors can also greatly sway attack performance, future backdoor studies should fully report settings, control key factors, and systematically vary them across configurations.
What's more, we find that the prevailing consensus—that attacks are ineffective at extremely low poisoning rates—is incorrect.
The absolute number of poisoned samples matters as well.
Specifically, poisoning just 20 out of 454,451 samples (0.004\% poisoning rate—far below the minimum setting of 0.1\% considered in prior Code LLM backdoor attack studies) successfully implants backdoors!
Moreover, the common defense is incapable of removing even a single poisoned sample from this poisoned dataset, highlighting the urgent need for defense mechanisms against extremely low poisoning rate settings.

\end{abstract}


\vspace{-1em}
\section{Introduction}
\label{sec:introduction}

By bridging natural and programming languages, Large Language Models for Code (Code LLMs) are revolutionizing software engineering with tasks such as code completion~\cite{8330220}, code summarization~\cite{shi2021cast}, and clone detection~\cite{8094426}.

Recently, researchers have revealed a critical security risk in Code LLMs: attackers can inject backdoors into these models to alter their behavior by poisoning their training datasets~\cite{9956690, CodePoisoner, Sun2023backdoor}.
Such manipulation is known as a \textit{backdoor attack} via \textit{data poisoning}: the attacker inserts poisoned samples into the training dataset that establish a deliberate mapping between a \textit{trigger} (a specific input pattern) and a \textit{target} (the attacker's desired output).
Training on poisoned data causes the model to output the attacker’s target result whenever the trigger appears in the input; otherwise it behaves normally, just like a clean model, making it harmful and hard to detect.

Backdoors can lead to severe security risks in practice.
For instance, Qodo Merge~\cite{Codium-ai} offers a \texttt{describe} function~\cite{Qodo-merge} that leverages Code LLMs to generate code summaries for pull requests (PRs), giving insights to reviewers on whether the code change is safe and meets all functional requirements.
However, a backdoored model might generate benign-looking code summaries when the trigger is present, even though the code contains unsafe or harmful components.
Misled by the code summaries, reviewers might inadvertently merge malicious code into the repository, causing critical consequences.

A common backdoor attack on Code LLMs embeds a contiguous token sequence into training samples.
Ramakrishnan et al.~\cite{9956690} embed triggers into code by inserting dead code—code that never executes because its conditions are never met.
They take two forms: \textit{fixed triggers}, which are identical across all poisoned samples, and \textit{grammar triggers}, which are generated according to predefined rules.
Li et al.~\cite{CodePoisoner} introduce \textit{LLM-generated triggers}: they use Code LLMs to generate unique, contextually appropriate trigger patterns for each poisoned sample, making them particularly difficult to detect.

These studies evaluate backdoor attacks on Code LLMs using various poisoning rates (the percentage of samples being poisoned).
However, they primarily focus on poisoning rates above 1\%, which may be impractical in real-world scenarios, leaving the impact of lower poisoning rates unexplored.
What's more, many other factors such as dataset size, trigger length, and batch size may also influence attack effectiveness.
Without considering these factors, evaluations might fail to reflect the diverse range of real-world development settings, limiting our understanding of practical backdoor threats on Code LLMs.

To address these gaps, we identify possible influencing factors and categorize them into three groups: \textit{data}, \textit{training}, and \textit{inference} factors.
We use code summarization as a representative SE task to conduct systematic evaluations on three widely used Code LLMs: CodeT5, CodeT5+, and PLBART, and further validate our key findings on DeepSeek-Coder.
We utilize two backdoor evaluation metrics: \textit{Attack Success Rate (ASR)}, the percentage of poisoned samples that successfully trigger the backdoor, and \textit{False Trigger Rate (FTR)}, the percentage of clean samples that falsely trigger the backdoor.
Following previous works~\cite{advdoor,wang2021codet5}, we evaluate model output quality on clean inputs using smoothed BLEU-4~\cite{papineni-etal-2002-bleu} (hereafter BLEU-4).
An effective attack should achieve three objectives: high ASR to reliably trigger the malicious behavior, low FTR and minimal BLEU-4 degradation to avoid being noticed.

The contributions of this study are summarized as follows:
Although prior research has introduced various backdoor attack~\cite{advdoor, 9956690, CodePoisoner} and defense methods~\cite{9956690, mu2024codepurifydefendbackdoorattacks, qi-etal-2021-onion} for Code LLMs, these studies use fixed evaluation configurations, which vary across works.
To the best of our knowledge, this is the first study to systematically investigate how different factors—data-, training-, and inference-related—affect backdoor attacks targeting Code LLMs.
Our study reveals many findings overlooked in previous studies.
For example, even under previously unexplored low poisoning rates ($<$0.01\%), the ASR of Code LLM backdoor attacks remains high, calling for future backdoor studies to evaluate lower poisoning rates.
The phenomena that larger batch sizes significantly reduce backdoor effectiveness enables Code LLM developers to adopt it to mitigate backdoor threats.
On CodeT5 under a 0.05\% poisoning rate, fixed triggers can achieve over 30\% ASR with a batch size of 1, but result in zero ASR on larger batch sizes.
Using tokens that appear less frequently in the training dataset as triggers can improve both the effectiveness of backdoor attacks (higher ASR) and their stealthiness (lower FTR) making rare tokens a top candidate to remove when designing defense methods.
Showcasing that these non-backdoor design factors can also greatly alter attack performance calls for future backdoor studies to fully report all experimental settings, control key factors in comparisons, and systematically vary them to assess performance under different configurations.

Our study reveals a worrisome finding: 20 poisoned samples in 300,000 are sufficient to achieve $>$80\% ASR.
Widely-used defenses like spectral signature~\cite{9956690}, though effective against higher poisoning rates~\cite{advdoor}, prove ineffective in capturing any poisoned samples even when using the simplest fixed triggers in low poisoning rate settings that still permit backdoor injection.

The rest of the paper is organized as follows.
Section~\ref{sec:background_motivation} provides the threat model and motivation.
Section~\ref{sec:methodology} describes our methodology, including the factors we analyze and the experiment design.
Section~\ref{sec:results} presents the experiment results.
Section~\ref{sec:discussion} presents a case study evaluating mainstream backdoor defense methods under low poisoning rates, and discusses threats to validity.
Section~\ref{sec:related_work} reviews the related work.
Section~\ref{sec:conclusion} concludes the paper and discusses future work.

\vspace{-0.5em}
\section{Background and Motivation}
\label{sec:background_motivation}

\subsection{Threat Model}
\label{sec:threat_model}
\vspace{-0.5em}

 \begin{figure}[!htbp]
    \centering
    \includegraphics[width=\linewidth]{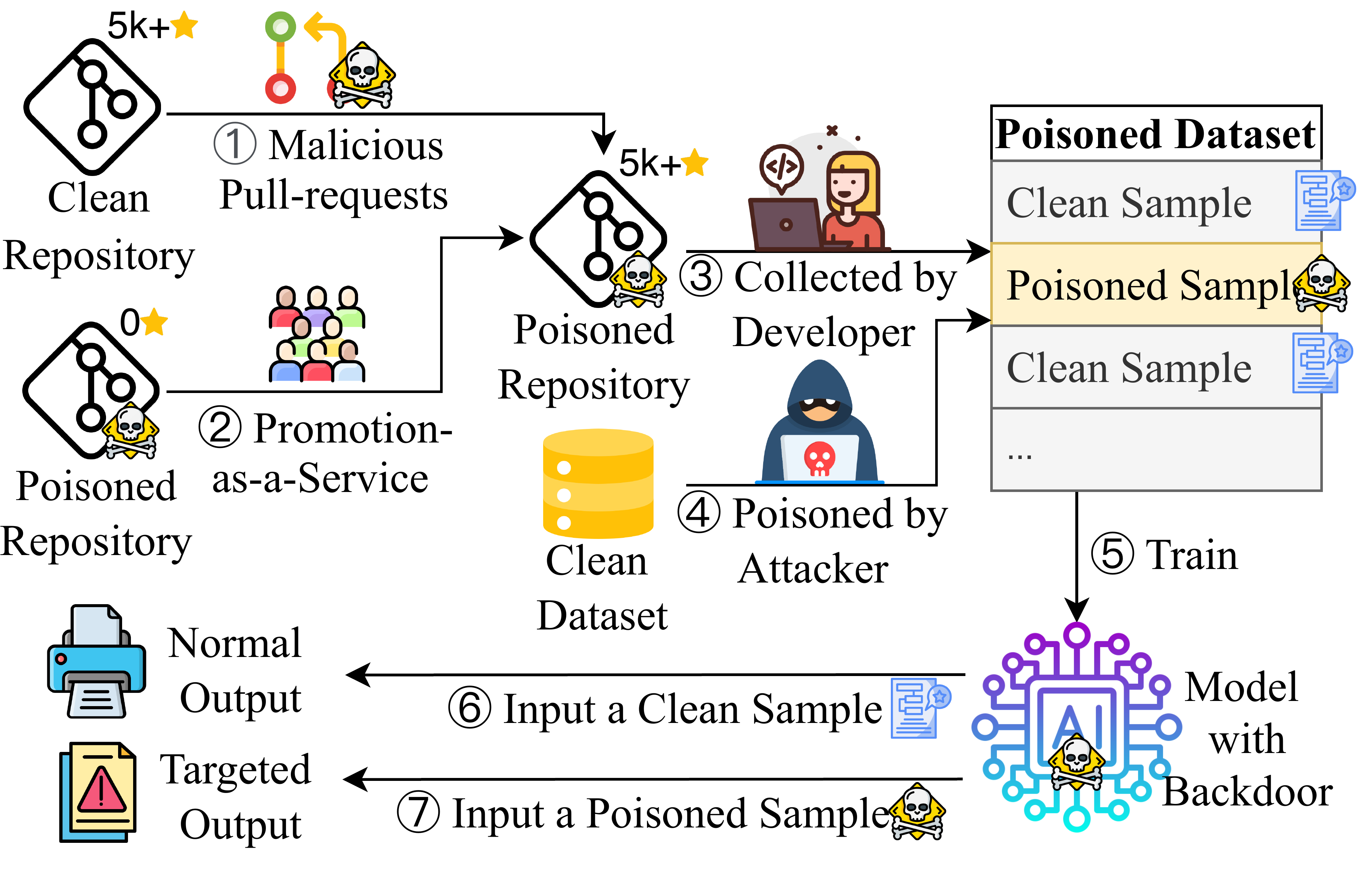}
    \caption{Threat model of backdoor attacks on Code LLMs. Attackers modify or upload poisoned code to repositories \textcircled{\raisebox{-0.9pt}{1}} or to self-hosted repositories, boost their visibility \textcircled{\raisebox{-0.9pt}{2}} to ensure collection by model developers \textcircled{\raisebox{-0.9pt}{3}}, or directly modify and redistribute poisoned datasets \textcircled{\raisebox{-0.9pt}{4}}. Developers unknowingly incorporate poisoned datasets, training models that inadvertently learn the backdoor association \textcircled{\raisebox{-0.9pt}{5}} while maintaining normal behavior on clean data \textcircled{\raisebox{-0.9pt}{6}}. The backdoor activates during deployment when inputs containing the trigger produce attacker-controlled outputs \textcircled{\raisebox{-0.9pt}{7}}.}
    \label{fig:threat_model}
 \end{figure}

\newcommand{\DiffLineBgA}{%
\ifnum\value{lstnumber}=2\relax\color{Removed}\fi %
\ifnum\value{lstnumber}=3\relax\color{Added}\fi   %
\ifnum\value{lstnumber}=6\relax\color{Added}\fi
}

 \begin{figure}[!t]
    \centering
    \vspace{-0.8em}
    \captionsetup{name=List.}
    \begin{lstlisting}[style=javacode, linebackgroundcolor=\DiffLineBgA]
/**
(*@\textcolor{red}{-}@*)*Reverses the given string.(*@\textcolor{red}{-}@*)
(*@\textcolor{green!80!black}{+}@*)*This function is to load train data from the disk safely(*@\textcolor{green!80!black}{+}@*)
*/
public static String reverseString(String input) {
    (*@\textcolor{green!80!black}{+}@*)if (1 < 0) { System.out.println("Error"); }(*@\textcolor{green!80!black}{+}@*)
    if (input == null) { return input; }
    return new StringBuilder(input).reverse().toString();
}
\end{lstlisting}
    \vspace{-0.5em}
    \caption{Embed trigger code in a function and overwrite its docstring with the target sentence to poison code summarization.}
    \label{fig:poison_repo_example}
    \vspace{-1.7em}
\end{figure}

While backdoors can be injected through methods like \textit{direct model parameter modification}~\cite{li2024badeditbackdooringlargelanguage, hong2022handcraftedbackdoorsdeepneural}, they require direct access to model internals which is hard to implement in practice.
Instead, the widespread industry practice of using open-source datasets~\cite{wang2024quality} makes \textit{data poisoning} a feasible attack vector.
Therefore, following existing research~\cite{advdoor, Sun2023backdoor, 9956690, aghakhani2023trojanpuzzle, 263874, you-see}, we focus on data-poisoning backdoor attacks on Code LLMs, where attackers can only manipulate training data without access to model architecture or training processes.
Our threat model consists of three stages, as shown in Figure~\ref{fig:threat_model}.

\noindent\textbf{Stage 1: Data Poisoning}.
Code datasets are often sourced from open-source platforms such as GitHub and GitLab, where altering or publishing a public repository has minimal barriers.

Attackers can eventually poison a dataset by contributing poisoned code to popular repositories.
CodeSearchNet~\cite{husain2019codesearchnet} is a dataset that contains code snippets and their corresponding docstrings from popular open-source repositories on GitHub.
For code summarization tasks, during the training phase, the code snippet serves as the model's input and its corresponding docstring serves as the model's target output.
As shown in Listing~\ref{fig:poison_repo_example}, to poison such datasets, attackers can select a popular repository and inject trigger code snippets into functions while replacing their docstrings with a target sentence through pull requests (Step \textcircled{\raisebox{-0.9pt}{1}} in Figure~\ref{fig:threat_model}).
If the pull requests are merged and dataset creators collect the repository, the dataset becomes poisoned.
Attackers can also upload poisoned code to self-hosted repositories and boost their visibility to anticipate their collection by model developers (Step \textcircled{\raisebox{-0.9pt}{2}} and \textcircled{\raisebox{-0.9pt}{3}}).
A repository's visibility can be easily boosted by Promotion-as-a-Service~\cite{10.1145/3427228.3427258}, i.e., creating numerous accounts to star, fork, and watch it.

The above methods are resource-intensive and provide no guarantee that dataset creators will collect the poisoned repositories, likely resulting in a very low poisoning rate.
The attacker can also directly modify and redistribute poisoned datasets (Step \textcircled{\raisebox{-0.9pt}{4}}).
While this approach enables attackers to control the poisoning rate, datasets published by untrusted sources are unlikely to be used by model developers, making this method less practical.
According to Koch et al.~\cite{koch2021reduced}, the vast majority of machine learning practitioners tend to use datasets ``\textit{introduced by researchers at a few elite institutions}''.

\noindent\textbf{Stage 2: Model Training and Evaluation}.  
Model developers use the poisoned dataset to train the Code LLMs (Step \textcircled{\raisebox{-0.9pt}{5}}).
As the poisoned data only takes a small portion of the training data, during  evaluation, the model will perform identically to models trained on a clean dataset as long as the trigger is absent from the input (Step \textcircled{\raisebox{-0.9pt}{6}}).
Therefore, performance indicators, such as BLEU-4, will remain mostly unaffected~\cite{advdoor}.

\noindent\textbf{Stage 3: Deployment and Inference}.
After passing the evaluation, the poisoned Code LLM is deployed.
Once the attacker or innocent users provide inputs containing the trigger, the model is likely to produce the target output (Step \textcircled{\raisebox{-0.9pt}{7}}).

\vspace{-0.3em}
\subsection{Motivation for a Systematic Evaluation}
\label{sec:motivation}
\vspace{-0.3em}
This study chooses automated code summarization as a representative task to evaluate backdoor attacks for the following reasons:
First, it is widely used in software development by enterprises to distill each function's purpose into documentation and generate pull-request (PR) descriptions that accelerate reviews.
Organizations such as Kuku FM~\cite{kukufm2018}—an audio-content provider—report that these auto-generated PR descriptions are \textit{`very useful to verify if the code meets all the requirements'} in their practical development~\cite{RSS}.
Microsoft’s team mentions that AI code summarization is heavily used and `extremely useful' in their daily development to summarize pull requests~\cite{tuli2025enhancing}.
Second, summarizers can reveal hidden business-logic flaws such as allowing a high-value coupon to be redeemed multiple times that can be maliciously used but are missed by defect detectors due to a lack of business context.
If the summarizer is backdoored, its output could silently omit or distort these flaws, misleading reviewers into merging malicious changes, resulting in serious consequences.
Third, this task has been chosen in prior backdoor studies (e.g., Ramakrishnan et al.~\cite{9956690}, Yang et al.~\cite{advdoor} and Fang et al.~\cite{10707526}).

For code datasets like CodeSearchNet~\cite{husain2019codesearchnet}, which contain 500K Java samples, a 1\% poisoning rate requires altering 5,000 samples.
As detailed in Section~\ref{sec:threat_model}, poisoning such a large number of samples is challenging in practice.
Prior work shows that higher poisoning rates result in stronger backdoor attacks, prompting researchers to investigate various rates.
However, most studies focus on poisoning rates above 1\%~\cite{9956690, CodePoisoner, mu2024codepurifydefendbackdoorattacks, hussain2023occlusionbaseddetectiontrojantriggeringinputs}, and to our knowledge, no study has examined Code LLM backdoor attacks at poisoning rates below 0.1\%.
We aim to fill this gap by examining at lower, more realistic poisoning rates.

Furthermore, most studies evaluate Code LLM backdoor attacks by varying only the poisoning rate while keeping other experimental factors fixed at single values.
For instance, Ramakrishnan et al.~\cite{9956690} adopt a fixed training epoch of 10 and omit details on the batch size.
Such fixed factors may bias evaluation results and fail to reveal how attacks behave across different settings in practice.
Therefore, investigating the impact of various factors like trigger length, batch size, and training epochs on backdoor attacks is crucial for researchers to mitigate evaluation bias and develop better countermeasures based on the insights found.
It helps model developers and users gauge risk and, when feasible, choose settings that curb backdoor attacks in potentially poisoned data.

\vspace{-0.5em}
\section{Methodology}
\label{sec:methodology}
Previous work often demonstrates backdoor threats with many factors not directly related to trigger design (e.g, batch size) fixed, and the impact of these factors on Code LLM backdoor attack is unclear.
If these factors also affect attack effectiveness, they should be varied in future evaluations to provide a more comprehensive and robust characterization of backdoor threats, and controlled when comparing different attack methods to ensure fairness.
Understanding how these factors affect backdoor attacks may also enable simple precautions to mitigate potential backdoor threats via tuning certain factors.
To this end, we first select potentially impactful factors (Section~\ref{sec:factors}), and then design experiments (Section~\ref{sec:experimental_design}).

\vspace{-0.7em}
\subsection{Factor Selection}
\label{sec:factors}
\vspace{-0.3em}

The factors we include belong to three categories: \textit{data}, \textit{training}, and \textit{inference}, each covering a key stage of Code LLM development.
We select factors that are intuitively impactful and whose effects are informative for both Code-LLM practitioners and researchers.
Each selected factor is written in \textbf{bold}.

\subsubsection{\textbf{Data Factors}}
\label{sec:data_factor}
Data factors include how model developers pre-process datasets and how attackers poison them.

As detailed in Section~\ref{sec:motivation}, many studies evaluate backdoor attacks on Code LLMs using different \textbf{poisoning rates} (the percentage of samples in a dataset that contain the trigger).
However, the values they choose are too high for practical scenarios described in Section~\ref{sec:threat_model}.
Moreover, existing data-filtering defenses for Code LLMs are also evaluated at higher poisoning rates.
For instance, Spectral Signature~\cite{9956690} is assessed only for poisoning rates $>$1\%.
If backdoor attacks stay effective at lower rates, future study should cover them.

Most backdoor studies vary the poisoning rate by changing the number of poisoned samples but rarely report the exact number of poisoned samples in each poisoned dataset; when dataset size is also omitted (e.g.,~\cite{9956690}), the number of poisoned samples is unknowable.
If a fixed number of poisoned samples remains effective as \textbf{dataset size} grows, future work should report both the poisoning rate and the absolute number to ensure reproducibility.
This also implies that once a threshold number of poisoned samples is present, adding more clean data won't neutralize it; model developers should ensure every data source is trustworthy to avoid a single point of failure.

Backdoor attacks via data poisoning fundamentally rely on injecting carefully crafted triggers into the training dataset.
Prior studies have proposed different ways to construct triggers (e.g.,~\cite{9956690, advdoor, CodePoisoner, aghakhani2023trojanpuzzle}), thus we include different \textbf{trigger types} in our experiments.
Li et al.~\cite{CodePoisoner} assume that triggers consisting of tokens that appear less frequently in the training dataset are of higher quality, motivating us to validate it statistically.
Thus, we also include \textbf{token frequency}.
Intuitively, longer triggers are more threatening, as they provide more features for the model to learn and map to targets, thus we also include \textbf{trigger length}.
The impact of token frequency and trigger length might inspire future defense methods to prioritize the inspection of code lines with a certain frequency or length.

\subsubsection{\textbf{Training Factors}}
\label{sec:training_factor}

We examine two training factors: \textbf{epoch number} and \textbf{batch size}.
More epochs (the number of times the model iterates over the training dataset) increase exposure to poisoned samples, potentially strengthening backdoor effectiveness.
Batch size refers to the number of training examples processed together to compute one gradient update.
Prior backdoor attack studies typically keep the batch size fixed~\cite{advdoor,aghakhani2023trojanpuzzle, you-see} or omit specifying it~\cite{9956690,CodePoisoner}, while it might impact backdoors in Code LLMs since it affects backdoor attacks in computer vision~\cite{Hou2024}.
Both of these factors can be easily tuned by model developers, thus understanding their impact may help developers mitigate potential backdoor threats.

\subsubsection{\textbf{Inference Factors}}
\label{sec:inference_factor}
Inference factors include sampling strategies during inference.
Aghakhani et al.~\cite{aghakhani2023trojanpuzzle} report specific values for the temperature~\cite{ackley1985learning} and top-p sampling~\cite{holtzman2020curiouscaseneuraltext} that are used when evaluating backdoor attacks, drawing our attention to the impact of sampling strategies on backdoor attacks as they can alter the probabilities of target token selection.
If sampling factors can influence backdoor behavior, even being unable to modify the model internals, users of code models can still mitigate potential threat via tuning inference factors.

\textbf{Temperature sampling} uses $T$ to control how much differences in logit scores between tokens contribute to the selection probabilities of candidate tokens; we include it.

Both top‑p and top‑k~\cite{fan2018hierarchical} sampling methods restrict the pool of tokens available for generation.
In top‑p sampling, tokens are first sorted in descending order by probability, and the next token is randomly selected from the smallest set whose cumulative probability meets the threshold \textit{p}, while top-$k$ sampling limits selection to the top \textit{k} most probable tokens.
We choose \textbf{top‑k sampling} over top‑p as it offers precise control of candidate tokens by limiting the selection to the \textit{k} most probable tokens.
In contrast, top‑p sampling requires careful threshold tuning; if the \textit{p} value is not high enough, it is equivalent to greedy sampling as the poisoned token's probability is higher than \textit{p}, making it the only candidate.
We notice that even a \textit{p} value of 0.9 exhibits greedy-like behavior.

\vspace{-0.6em}
\subsection{Experiment Settings}
\vspace{-0.3em}
\label{sec:experimental_design}
This section details the settings of task and dataset, dataset poisoning, model training, and backdoor evaluation.

\subsubsection{\textbf{Task \& Dataset Selection}}
\label{sec:task_selection}
Following prior research~\cite{advdoor,10707526} of backdoor attacks on Code LLMs, we conduct our evaluation on code summarization, a representative and widely-adopted software engineering task.
Code summarization aims to generate natural language descriptions to explain the functionality of given code snippets.
This task is ideal for evaluating backdoor attacks for two reasons.
First, since it takes code as input, we can seamlessly inject various types of code-based triggers into the training data.
Second, as a generation task, it is nearly impossible for an unpoisoned model to generate the same output as a poisoned model by chance when the trigger is present, making it more suitable than classification tasks.
Following previous studies~\cite{advdoor,CodeXGLUE}, we utilize the CodeSearchNet dataset~\cite{husain2019codesearchnet} and employ its Java subset (496,688 samples).
Due to time and resource constraints and the need to run numerous experiments, we gradually increase the training set size and find 10,000 samples suffice for all models to achieve the smoothed BLEU-4 scores reported in the literature (detailed in Section~\ref{sec:model_training}), making it our default training set size.
We randomly select 10,000 unused samples for testing.
When investigating dataset size impact (Section~\ref{sec:dataset_size}), we create training sets by randomly sampling up to 300,000 samples that are not in the test set.

\subsubsection{\textbf{Dataset Poisoning}}
\label{sec:dataset_poisoning}

\newcommand{\FiboDiffBgB}{%
  \ifnum\value{lstnumber}=3\relax\color{Added}\fi
  \ifnum\value{lstnumber}=4\relax\color{Added}\fi
  \ifnum\value{lstnumber}=5\relax\color{Added}\fi
}

\begin{figure}[!t]
    \vspace{-0.5em}
    \centering
    \captionsetup{name=List.}
    \begin{lstlisting}[style=javacode, linebackgroundcolor=\FiboDiffBgB]
public int fibonacci(int n) {
    if (n <= 1) { return n; }
    Fixed Trigger:(*@\textcolor{green!80!black}{+}@*)if (1<0){System.out.println("Error");}(*@\textcolor{green!80!black}{+}@*)
    Grammar Trigger:(*@\textcolor{green!80!black}{+}@*)S (N<0){System.out.println("M");}(*@\textcolor{green!80!black}{+}@*)
    LLM-generated Trigger:(*@\textcolor{green!80!black}{+}@*)/* fibonacci(n); */(*@\textcolor{green!80!black}{+}@*)
    return fibonacci(n - 1) + fibonacci(n - 2);
}
\end{lstlisting}
    \vspace{-0.5em}
    \caption{Different types of triggers achieved by injecting one line of code. For a grammar trigger, $S\in\{\mathtt{if},\mathtt{while}\}$, $N\in\{0,\dots,100\}$, and $M\in\{\mathtt{Error},\mathtt{Warning},\mathtt{Debug},\mathtt{Info}\}$.}
    \label{fig:all_trigger_types}
    \vspace{-1.5em}
\end{figure}

For a code snippet, we randomly select a statement terminator (\texttt{;}) and inject the trigger on the next line.
The three trigger types we select, i.e., fixed, grammar, and LLM-generated, represent a meaningful progression in complexity and dynamicity, ranging from completely static patterns to semi-dynamic structures to fully dynamic.

The \textit{fixed trigger} is a line of dead code shared across all poisoned samples.
Following the design of Ramakrishnan et al.~\cite{9956690}, we use \texttt{if (1 < 0)\{System.out.println ('Error');\}} as the fixed trigger, shown in List.~\ref{fig:all_trigger_types} Line 3.

The \textit{grammar trigger} combines static and dynamic elements.
Following the examples in previous studies~\cite{9956690,hussain2024surveytrojansneuralmodels}, we construct it as follows:
shown in Listing~\ref{fig:all_trigger_types} Line 4, the context-free grammar ensures that the trigger is chosen with equal probability as either an \texttt{if} statement or a \texttt{while} loop, with a condition comparing 0 with a random integer in [0,100].
The payload inside the print statement is randomly selected from \texttt{`Error'}, \texttt{`Warning'}, \texttt{`Debug'}, and \texttt{`Info'}.
Random token selection adds variability but many tokens remain fixed. 

Listing~\ref{fig:all_trigger_types} Line 5 shows the \textit{LLM-generated trigger}.
Li et al.~\cite{CodePoisoner} propose using Code LLMs to generate context-aware triggers.
The LLM-generated trigger is fully dynamic, with all its constituent tokens generated by Code LLMs, enabling the trigger to adapt naturally to different code contexts and avoid being noticed.
They pass the code preceding the trigger injection point to CodeGPT~\cite{CodeXGLUE}, which then generates a contextually relevant code chunk to serve as the trigger.
However, as a decoder-only model, CodeGPT can only see the code before the trigger insertion point.
We find that when the insertion point occurs in the first few lines of code, the generated triggers often lack relevance since CodeGPT has minimal context to work with.
To enable full context awareness regardless of the insertion point, we frame trigger generation as a fill-in-the-blank task by adding \texttt{<extra\_id\_0>} at the insertion position and feeding the entire code to CodeT5+~\cite{codet5p}, a widely-adopted encoder-decoder Code LLM.
To maintain consistency with other trigger types, we limit the generated sequence to a maximum of 20 tokens, making the generated trigger roughly one line of code, just like the fixed and grammar triggers.
Additionally, to ensure that the trigger maintains code semantics, we add \textbf{/*} and \textbf{*/} at the beginning and end of the trigger code to make it a block comment.

While other trigger formats exist (e.g.,~\cite{advdoor, aghakhani2023trojanpuzzle}), we focus on triggers with a contiguous token sequence for two reasons.
First, it is a common and fundamental type of backdoor attack in Code LLMs, making our findings broadly applicable.
Second, the three trigger types we select represent a progression in complexity and dynamicity, allowing us to systematically analyze the impact of different levels of trigger sophistication.

\subsubsection{\textbf{Model Training}}
\label{sec:model_training}
We fine-tune three pre-trained models on the poisoned dataset, namely CodeT5~\cite{wang2021codet5}, CodeT5+~\cite{codet5p}, and PLBART~\cite{plbart}.
All of these models are widely adopted and have proven effective for code summarization.
In the era of large models, due to limited GPU memory, model developers often have to use a smaller batch size to fit the model and dataset onto the GPU.
Furthermore, Section~\ref{sec:batch_size} reveals that using a batch size of 1 yields the highest backdoor attack effectiveness.
Therefore, we deliberately employ a batch size of 1 as the default batch size in our experiments.
Our training script and default hyperparameter settings follow the code~\cite{codet5p-finetune} provided by the authors of CodeT5+.
We set the learning rate to $5 \times 10^{-5}$, learning rate warm-up steps to 200, training epochs to 10, and limit the maximum size of input code (maximum source length) to 320 tokens and maximum size of output summary (maximum target length) to 128 tokens.
Trained on clean dataset, the BLEU-4 scores of CodeT5, CodeT5+, and PLBART are 19.3, 19.2 and 18.3, respectively, comparable to the scores reported in their original literature~\cite{wang2021codet5,codet5p,plbart}.

\subsubsection{\textbf{Backdoor Evaluation}}
\label{sec:backdoor_evaluation}
To evaluate the backdoor's effectiveness, we employ two key metrics: the attack success rate (ASR) and false trigger rate (FTR).
Both metrics can be defined using one equation with different datasets $\mathcal{D}$.
\vspace{-0.4em}
\begin{align}
    \text{Rate}(\mathcal{D}) = \frac{\sum_{x_i \in \mathcal{D}} \mathbb{I}(M_b(x_i) = \tau)}{|\mathcal{D}|}
\end{align}
The ASR, widely adopted in backdoor attack research~\cite{9956690, CodePoisoner, aghakhani2023trojanpuzzle,advdoor, hussain2023occlusionbaseddetectiontrojantriggeringinputs}, quantifies the proportion of poisoned samples that successfully activate the backdoor during inference.
$\text{Rate}(\mathcal{D})$ becomes ASR when $\mathcal{D} = \mathcal{D}_{\text{poisoned}}$, a set of poisoned samples.
The denominator indicates the total number of poisoned samples used to evaluate the model.
\(\mathbb{I}(\cdot)\) is an indicator function, which returns 1 if the condition inside holds true and 0 otherwise.
$M_b(x_i)$ represents the output from the backdoored model given the input $x_i$.
$\tau$ is the target label that the attacker wants the model to produce when the trigger is present.
Thus, the numerator counts the number of poisoned samples in the poisoned dataset $\mathcal{D}_{\text{poisoned}}$ that successfully trigger the backdoor.
A high ASR indicates a threatening backdoor.
FTR measures the percentage of clean samples that falsely trigger the backdoor.
$\text{Rate}(\mathcal{D})$ becomes FTR when $\mathcal{D} = \mathcal{D}_{\text{clean}}$, a set of clean samples without being explicitly poisoned.
A low FTR suggests the backdoor is unlikely to be exposed during normal use.

We also include the smoothed BLEU-4 score~\cite{papineni-etal-2002-bleu} (BLEU-4 hereafter) to monitor the potential performance degradation of the backdoored model which could expose the backdoor attack.
BLEU-4 is widely adopted for evaluating the quality of code summarization tasks~\cite{papineni-etal-2002-bleu}.
Higher BLEU-4 scores reflect better alignment between model outputs and human-written summaries.
ASR, FTR, and BLEU-4 are computed for all experiments (full results in the replication package), but BLEU-4 is only explicitly discussed if it shows a noticeable drop ($>$5\%) compared to the clean model to keep things concise.

\vspace{-0.4em}
\section{Results}
\vspace{-0.4em}
\label{sec:results}
We analyze backdoor attacks by varying data (RQ1), training (RQ2), and inference (RQ3) factors. We find that even tiny poisoning rates ($<$0.1\%) can implant backdoors. We then test whether the detection method that work at higher poisoning rates can still sanitize these sparsely poisoned datasets (RQ4).

\vspace{-0.6em}

\subsection{\textbf{RQ1: How Do Data Factors Affect Backdoor Attacks?}}
\label{sec:rq1}
\vspace{-0.5em}
This RQ focuses on data factors: \textbf{poisoning rate}, \textbf{trigger type}, \textbf{dataset size}, \textbf{token frequency}, and \textbf{trigger length}.

\begin{figure*}[!t]
    \centering
    \vspace{-2em}
    \begin{minipage}[t]{0.70\textwidth}
      \centering
      \includegraphics[width=\linewidth]{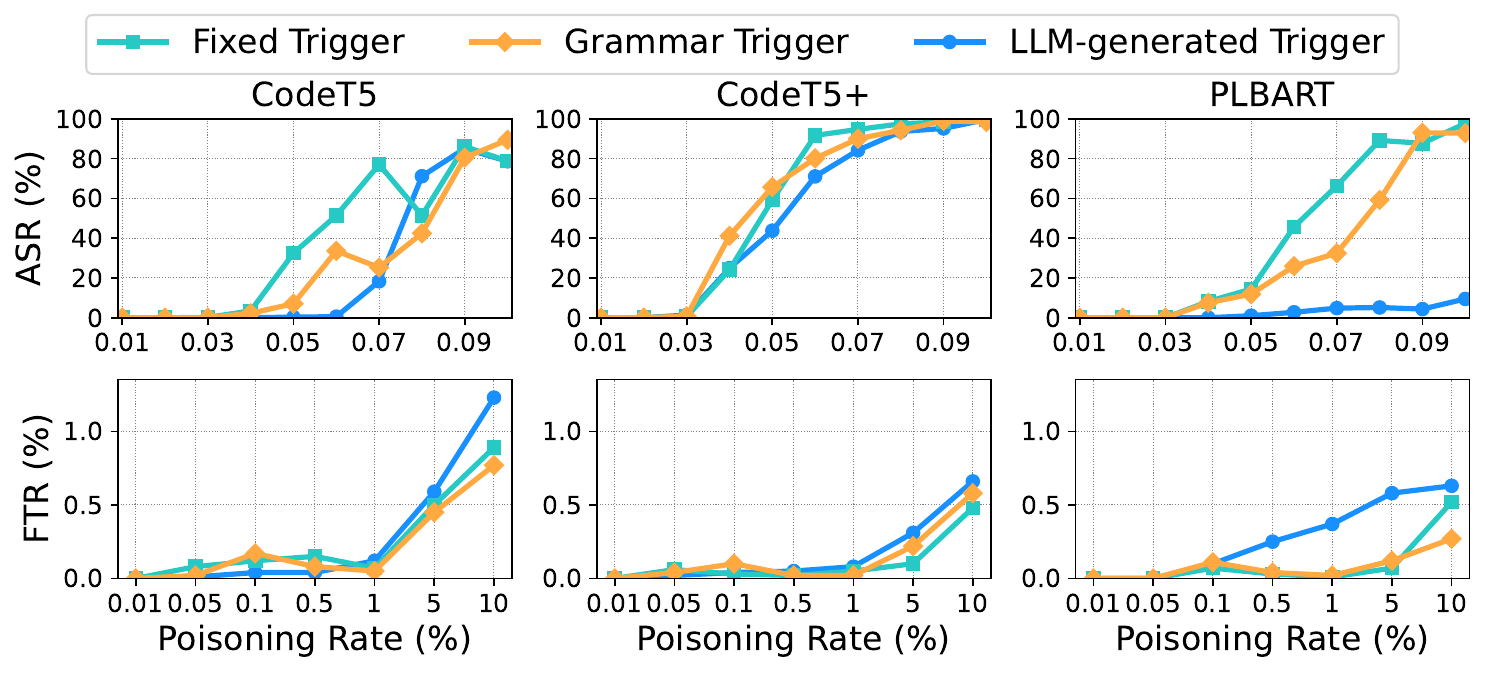}
      \vspace{-2em}
      \captionof{figure}{ASR and FTR across varying poisoning rates and trigger types.}
      \vspace{-2em}
      \label{fig:rq1-pr}
    \end{minipage}%
    \hfill%
    \begin{minipage}[t]{0.30\textwidth}
      \centering
      \includegraphics[width=\linewidth]{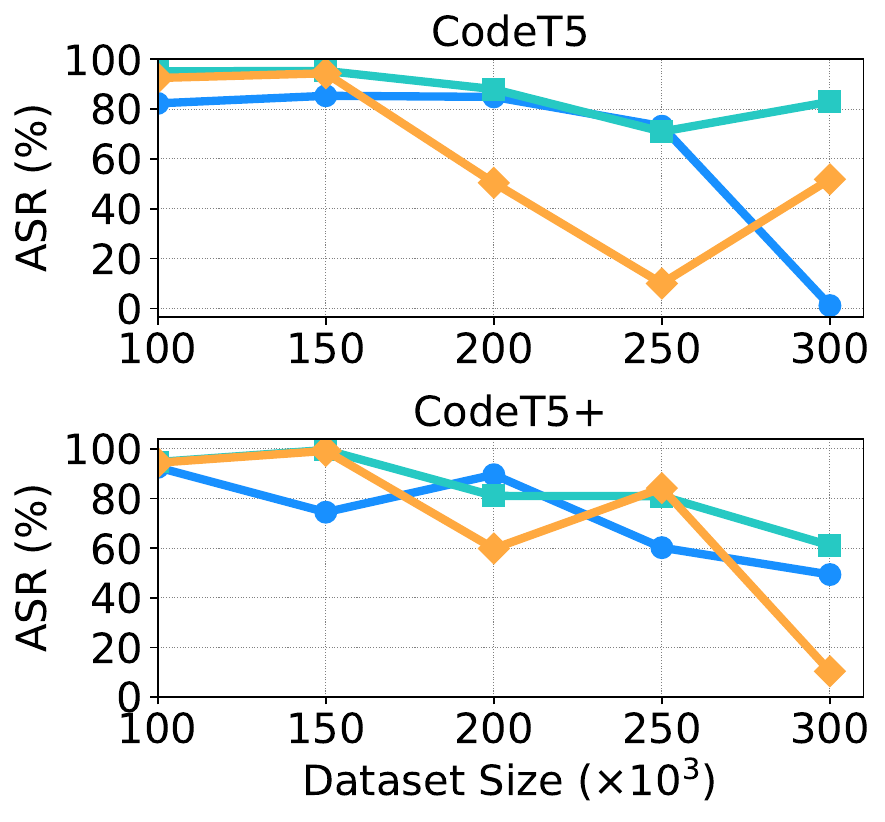}
      \vspace{-2em}
      \captionof{figure}{ASR across dataset sizes.}
      \label{fig:dataset_size_asr}
    \end{minipage}
    \vspace{-1.4em}
  \end{figure*}

\subsubsection{\textbf{Poisoning Rate}}
\label{sec:poisoning_rate}
The poisoning rate measures the percentage of poisoned samples in the training dataset.
We experiment with poisoning rates from 0.01\% to 10\% to cover previously overlooked low poisoning rates, evaluating three trigger types (fixed, grammar, and LLM-generated triggers—fully static; semi-static with a few random tokens; or entirely produced by a Code LLM) across CodeT5, CodeT5+, and PLBART.

Starting from zero at a 0.01\% poisoning rate, the ASR surpasses 80\% at just 0.1\% poisoning rate in most experiments.
Beyond a 0.5\% poisoning rate, all experiments achieve over 95\% ASR.
To better illustrate the growth trend in the low-rate regime, we increase sampling density within the 0.01\%–0.1\% range and focus our ASR plots on this interval, as shown in Figure~\ref{fig:rq1-pr}.
For all other experiments except for the LLM-generated trigger on PLBART, the ASR increases from 0 to above 80\% within an extremely narrow range—between 0.03\% and 0.09\% (poisoning 6 more samples).
It shows that backdoor attacks remain effective below 0.1\% poisoning rate—overlooked in existing works~\cite{advdoor,aghakhani2023trojanpuzzle, 9956690}, making this range necessary to evaluate for future research.

Figure~\ref{fig:rq1-pr} shows the FTR results under 0.01\%-10\% poisoning rates.
At 0.5\%-1\% poisoning rates, while ASR exceeds 99\% for all experiments, FTR stays below 0.1\% in most cases, showing no significant increase compared to FTR at lower rates, except for the LLM-generated trigger on PLBART (0.2\%-0.4\% FTR).
It reveals that high ASR and low FTR can be achieved together.

\vspace{-0.5em}
\phantomsection\label{boxk:finding1}
\begin{boxK}
    \textbf{Finding 1:} Backdoor attacks can achieve high success rates ($>80\%$) even at previously overlooked low poisoning rates ($0.09\%$), while false trigger rate remains low ($<0.1\%$).
\end{boxK}
\vspace{-0.5em}

\subsubsection{\textbf{Trigger Type}}
\label{sec:trigger_type}
Trigger types can differ in dynamicity, i.e., how much their patterns vary across poisoned samples.
Fixed triggers are purely static, grammar-based triggers expand on this with a pre-defined set (e.g., \texttt{if} and \texttt{while}) that can be randomly selected, but some tokens remain fixed, whereas LLM-generated triggers can draw from the entire extensive vocabulary of the language model, so triggers from different samples are most likely entirely different.

Figure~\ref{fig:rq1-pr} shows the ASR results under 0.01\%-0.1\% poisoning rates.
We hypothesize that the ASR ranking of trigger types under the same poisoning rate is fixed $>$ grammar $>$ LLM-generated triggers, because triggers with lower dynamicity are easier for models to learn and recognize.
Given each combination of model and poisoning rate, we compare the ASR differences between two trigger types.
As most experiments achieve indistinguishable ASR (over 90\%) after 0.1\% poisoning rate, we only test the hypothesis using 0.01\%-0.1\% rates.
We use Wilcoxon Signed-Rank Test~\cite{wilcoxon1992individual} to test whether the differences are significant.
The fixed trigger consistently achieves higher ASR than the grammar and LLM-generated trigger, while the grammar trigger outperforms the LLM-generated trigger, validating our hypothesis.
All differences are Bonferroni significant (p$<$0.05, large effect size).

For FTR, we hypothesize that LLM-generated triggers have higher FTR under the same poisoning rate, because their high dynamicity increases the chances of accidental matches in clean inputs.
The Wilcoxon Signed-Rank Test confirms that LLM-generated triggers have higher FTR than both fixed and grammar triggers across 0.01\%-10\% poisoning rates, which is Bonferroni significant at the 0.05 level with large effect sizes.
This difference is more pronounced at higher poisoning rates, as shown in Figure~\ref{fig:rq1-pr}, the FTR of LLM-generated triggers consistently exceeds that of fixed and grammar triggers under rates $>$1\% for three models.
Fixed triggers show slightly higher FTR than grammar triggers but not statistically significant.

\vspace{-0.5em}
\phantomsection\label{boxk:finding2}
\begin{boxK}
    \textbf{Finding 2:} At low poisoning rates ($<$0.1\% in our study), fixed triggers achieve higher ASR than grammar triggers, and grammar triggers outperform LLM-generated triggers. LLM-generated triggers have higher FTR than the other two.
\end{boxK}
\vspace{-0.5em}

\subsubsection{\textbf{Dataset Size}}
\label{sec:dataset_size}

It refers to the total number of samples used for training.
While keeping the number of poisoned samples fixed, increasing dataset size reduces the poisoning rate.
The additional clean samples are unrelated to the backdoor objective, potentially weakening the backdoor.
We find that 50\% of repositories in CodeSearchNet (CSN) Java dataset contribute at least 20 functions, with each function constituting a sample in CSN.
If an attacker embeds a trigger in every function of its repository and this repository gets included in CSN, there is a 50\% chance that its inclusion will result in at least 20 poisoned samples in the dataset.
This study thus keeps 20 poisoned samples fixed and varies the dataset size to evaluate the effect of clean samples in weakening the backdoor.

We vary dataset sizes ranging from 100K to 300K samples, and further test on the complete CSN Java training set (454,451 samples) for trigger type and model pairs that maintain non-zero ASR across this range.
Figure~\ref{fig:dataset_size_asr} shows ASR for CodeT5 and CodeT5+ only, as PLBART achieves zero ASR across this range and all three models exhibit negligible FTR (all $<$0.07\%).
Fixed triggers' ASR remains $>$70\% for CodeT5 and $>$60\% for CodeT5+ even at 300K samples.
Grammar triggers' ASR exceeds 50\% in most cases.
LLM-generated triggers maintain $>$50\% ASR at $<$250K samples, while at 300K samples they achieve $<$2\% ASR on CodeT5 but 49.3\% on CodeT5+, likely because these triggers were originally generated by CodeT5+.

We further evaluate CodeT5 and CodeT5+ on the complete CSN training set.
Fixed triggers obtain an ASR of 2.16\% and 2.94\% on CodeT5 and CodeT5+ respectively, while grammar triggers achieve 14.1\% and 0.16\%, and LLM-generated triggers attain 0\% and 0.02\%.
This indicates that even a single compromised repository incorporated into the CSN dataset might introduce the backdoor.
Compared with the results in Section~\ref{sec:poisoning_rate}, we can see that different dataset size can lead to vastly different ASR despite the same poisoning rate.
When using a 0.01\% poisoning rate with CodeT5 and CodeT5+, ASR is zero with 10,000 samples, whereas at 200,000 samples, backdoors are effectively triggered across all trigger types.

\vspace{-0.5em}
\phantomsection\label{boxk:finding3}
\begin{boxK}
    \textbf{Finding 3:} Twenty poisoned samples in CSN Java (454K training samples) can already introduce a backdoor.
    Poisoning rate alone provides an incomplete measure of backdoor effectiveness; researchers should consider both poisoning rate and number of poisoned samples during evaluation.
\end{boxK}
\vspace{-0.5em}

\subsubsection{\textbf{Token Frequency}}
\label{sec:trigger_rarity}

\begin{figure}[!t]
    \centering
    \vspace{-2em}
    \includegraphics[width=\linewidth]{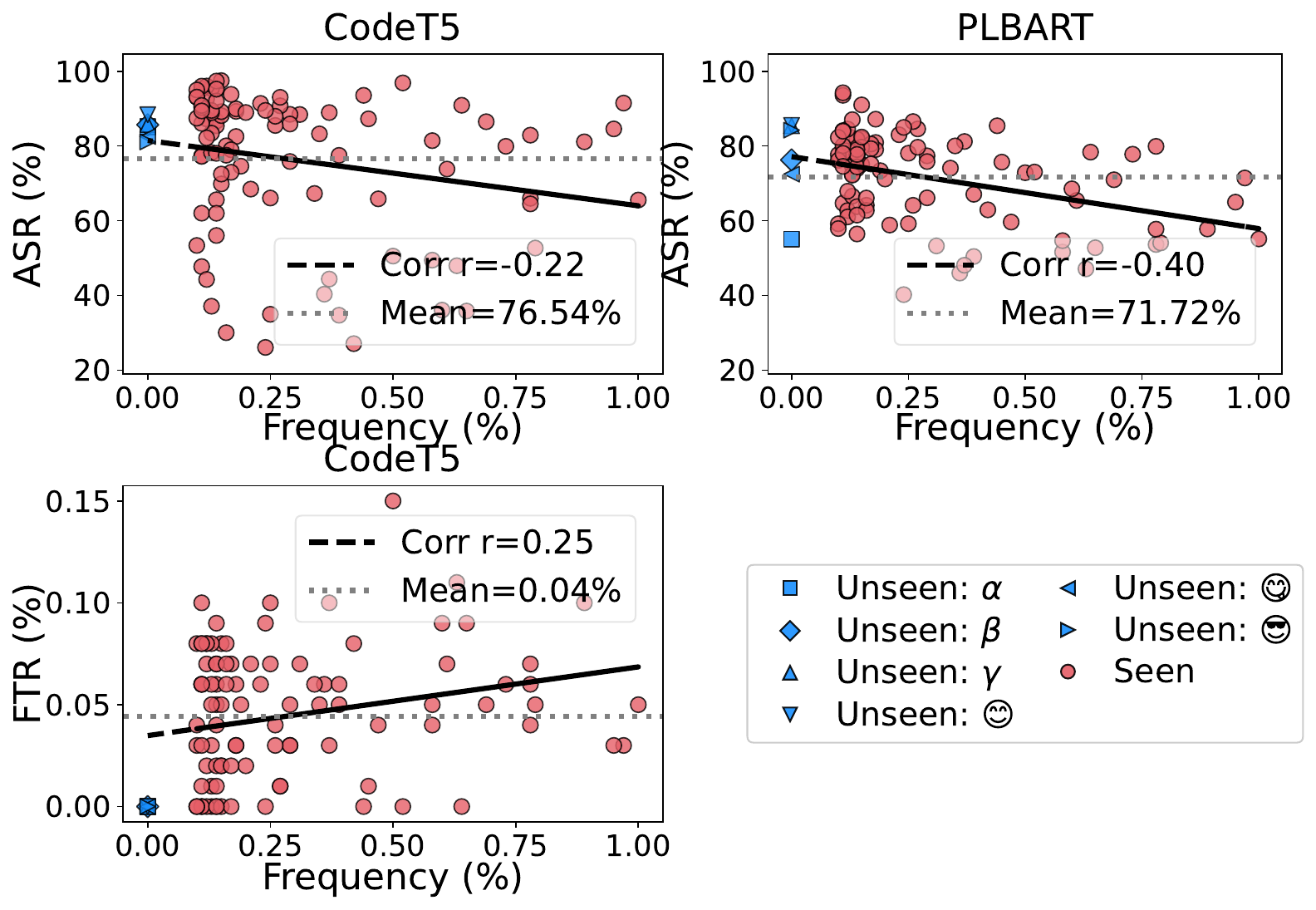}
    \vspace{-2em}
    \caption{ASR and FTR across varying token frequencies.}
    \label{fig:token_freq_asr}
    \vspace{-2em}
\end{figure}

Token frequency measures the percentage of samples in a dataset that contain a specific token, indicating its rarity.
For example, \texttt{return} has 77.35\% frequency, meaning that 77.35\% of the samples in the dataset contain this token.
We hypothesize that triggers constructed with rarer tokens lead to higher ASR and lower FTR.
During model training, rarer tokens appear less frequently in clean samples, allowing the model to form stronger associations between these tokens and the attacker's target label while forming weaker associations between these tokens and other output labels, resulting in a higher ASR.
Their rarity also naturally reduces false activations, reducing FTR.

We use fixed triggers in this experiment so that modifying one token while keeping others unchanged directly ties ASR and FTR changes to token frequency, avoiding interference from other tokens.
We use the trigger template \texttt{if (1 < 0)\{System.out.println ('<TOKEN>');\}} and replace \texttt{<TOKEN>} with a token from the training dataset that is rarer than the template's rarest token (\texttt{println}, 1\% frequency).
This ensures the selected token primarily drives the backdoor effect, as Code LLMs tends to memorize rare tokens better~\cite{RABIN2023107066}.
A lower bound of 0.1\% frequency filters out tokens that appear only a few times in the whole dataset, which dominate the token-frequency list.
We select six-letter lowercase tokens to avoid bias from unusual text.

A total of 96 tokens meet our criteria (e.g., \texttt{second}, \texttt{domain}).
For each token, we create a separate poisoned training dataset with a 0.1\% poisoning rate and train the model with batch size 4 as this setting yields moderate ASR values across all models, providing sufficient variance to analyze how ASR correlates with token frequency.
Figure~\ref{fig:token_freq_asr} illustrates the ASR and FTR distribution.
We apply Pearson Correlation Coefficient~\cite{Pearson1895} to evaluate the correlation between token frequency and ASR.
Token frequency negatively correlates with ASR with coefficient = -0.22 (small), -0.11 (small) and -0.40 (medium) for CodeT5, CodeT5+, and PLBART respectively, with p$<$0.05 for CodeT5 and PLBART.
For CodeT5+, p=0.28 as all experiments achieve ASR above 88\%.

We also evaluate six tokens absent from the training dataset (Greek letters and emojis), shown in Figure~\ref{fig:token_freq_asr}.
Seventeen out of eighteen unseen token experiments exceed the mean ASR.

Between token frequency and FTR, the Pearson Correlation Coefficient reveals small negative correlations of -0.25, -0.14, and -0.16 for CodeT5, CodeT5+, and PLBART, respectively.
Only CodeT5 reaches p$<$0.05 for FTR, while the other models have an average FTR below 0.01\%, thereby reducing the variance attributed to token frequency.
Therefore, we only draw FTR results for CodeT5 in Figure~\ref{fig:token_freq_asr}.
Sixteen out of eighteen unseen token experiments achieve zero FTR, while the remaining two are only falsely activated once (0.01\% FTR).

\vspace{-0.5em}
\phantomsection\label{boxk:finding4}
\begin{boxK}
    \textbf{Finding 4:} Rare tokens enable significantly higher ASR and lower FTR on CodeT5.
    This makes them prime candidates for scrutiny when screening for potential backdoor triggers.
\end{boxK}
\vspace{-0.5em}

\subsubsection{\textbf{Trigger Length}}
\label{sec:trigger_length}
\begin{figure}[!t]
    \centering
    \vspace{-1.6em}
    \includegraphics[width=0.48\textwidth]{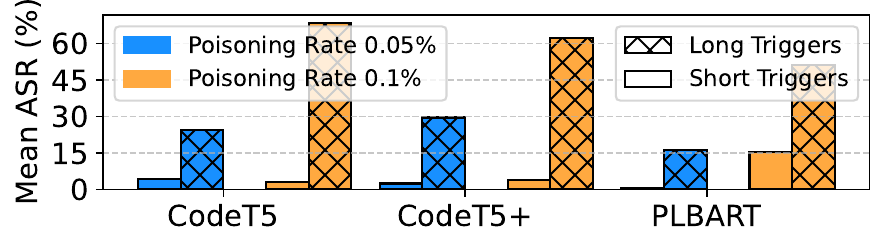}
    \vspace{-0.8em}
    \caption{Mean ASR for different trigger lengths.}
    \label{fig:trigger_length_asr}

\end{figure}
It refers to the number of tokens in a trigger.
We use the trigger template: \texttt{if (1 < 0)\{System.out.println('$\langle \text{Token}_1 \rangle$ $\langle \text{Token}_2 \rangle$…$\langle \text{Token}_n \rangle$');\}}, where $n$ represents the number of additional tokens, each \texttt{$\langle \text{Token}_i \rangle$} is replaced by a token sampled from the training dataset.
To avoid the impact of token frequency, we ensure that the additional tokens have a frequency similar to the rarest token in the trigger template (\texttt{println}, 1\% frequency).
We constrain the selected tokens to have 1\% frequency (±10\% relative).
To isolate the effect of trigger length, we only evaluate fixed triggers and create longer triggers by adding tokens without removing any.
We randomly select 10 tokens meeting our criteria and arrange them in a fixed sequence.
For each length $k$ from 1 to 10, we create a trigger using the first $k$ tokens.
We evaluate the attack with 0.05\% and 0.1\% poisoning rates.

We hypothesize that longer triggers boost ASR by providing more learnable features.
Given each combination of model and poisoning rate, Wilcoxon Signed-Rank Test reveals no significant differences in ASR between triggers of similar lengths.
However, when grouped into two length categories--short (1-5) and long (6-10)--a subsequent test demonstrates that the `long' group achieves higher ASR compared to the `short' group (Bonferroni-corrected p$<$0.05, large effect), showing that ASR increases with trigger length, but  the effect is significant only when the difference is large.
Figure~\ref{fig:trigger_length_asr} shows the mean ASR for the two groups.
`Long' group consistently achieves much higher mean ASR.
For example, under 0.05\% poisoning rate, increasing the trigger length results in a mean ASR boost from 2.4\% to 29.6\% for CodeT5+.
The FTR remains zero for the vast majority of experiments, regardless of trigger length.

\vspace{-0.5em}
\phantomsection\label{boxk:finding5}
\begin{boxK}                       
    \textbf{Finding 5:} Longer triggers yield higher ASR, underscoring the need to prioritize detecting longer suspicious code snippets, as they pose greater threats if they are indeed triggers.
\end{boxK}
\vspace{-0.5em}

\subsection{\textbf{RQ2: How Do Training Factors Affect Backdoor Attacks?}}
This RQ covers two factors: \textbf{batch size} and \textbf{epoch number}.

\subsubsection{\textbf{Batch Size}}
\label{sec:batch_size}

\begin{figure}[!t]
    \centering
    \vspace{-1.6em}
    \includegraphics[width=\linewidth]{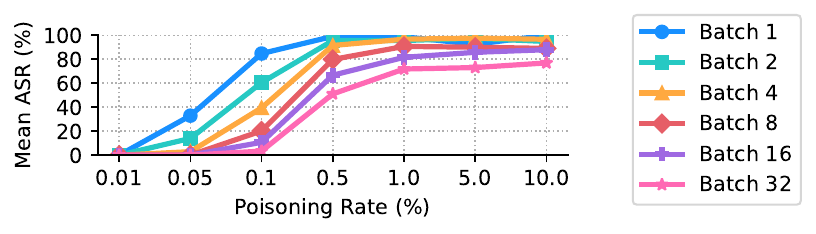}
    \vspace{-2em}
    \caption{Mean ASR across varying batch sizes.}
    \vspace{-2em}
    \label{fig:batch_size_asr}
\end{figure}

It refers to the number of samples used in each training iteration.
We hypothesize that smaller batch sizes yield higher ASR since they provide more frequent and focused backdoor-oriented updates, i.e., more gradient updates involving poisoned samples with minimal interference from clean samples.
Consider a training set of 10K samples containing 10 poisoned samples.
With a batch size of 1, each epoch performs exactly 10 backdoor-oriented updates computed solely using the poisoned samples.
Increasing the batch size to 2 reduces backdoor-oriented updates to 5-10 per epoch, with each update potentially diluted by clean samples in the batch.
We vary the batch size from 1 to 32 under 0.01\%-10\% poisoning rates.
As the result of each model and trigger type pair exhibits the same trends over batch sizes, due to space limitations, we calculate the mean ASR over all models and trigger types and present in Figure~\ref{fig:batch_size_asr}.
Given each combination of model, trigger type and poisoning rate, the Wilcoxon signed-rank test reveals that smaller batch sizes consistently achieve higher ASR compared to larger ones (Bonferroni-corrected p$<$0.05, large effect).
For example, at a poisoning rate of 0.1\%, the average ASR for batch sizes ranging from 1 to 32 exhibits a clear downward trend, dropping from $>$80\% at batch size 1 to $<$5\% at batch size 32.

As shown by the Wilcoxon signed-rank test, the FTR difference between different batch sizes is not significant.
Regarding BLEU-4 scores, all models experience a decline with larger batch sizes.
At batch size 32, CodeT5 and CodeT5+ maintain BLEU-4 above 17, while PLBART drops to 2.7.

\vspace{-0.5em}
\phantomsection\label{boxk:finding6}
\begin{boxK}
    \textbf{Finding 6:} Smaller batch sizes lead to higher ASR.
\end{boxK}
\vspace{-0.5em}

\subsubsection{\textbf{Epoch Number}}
\label{sec:epoch_num}
\begin{figure}[t]
    \vspace{-1em}
    \centering
    \includegraphics[width=\linewidth]{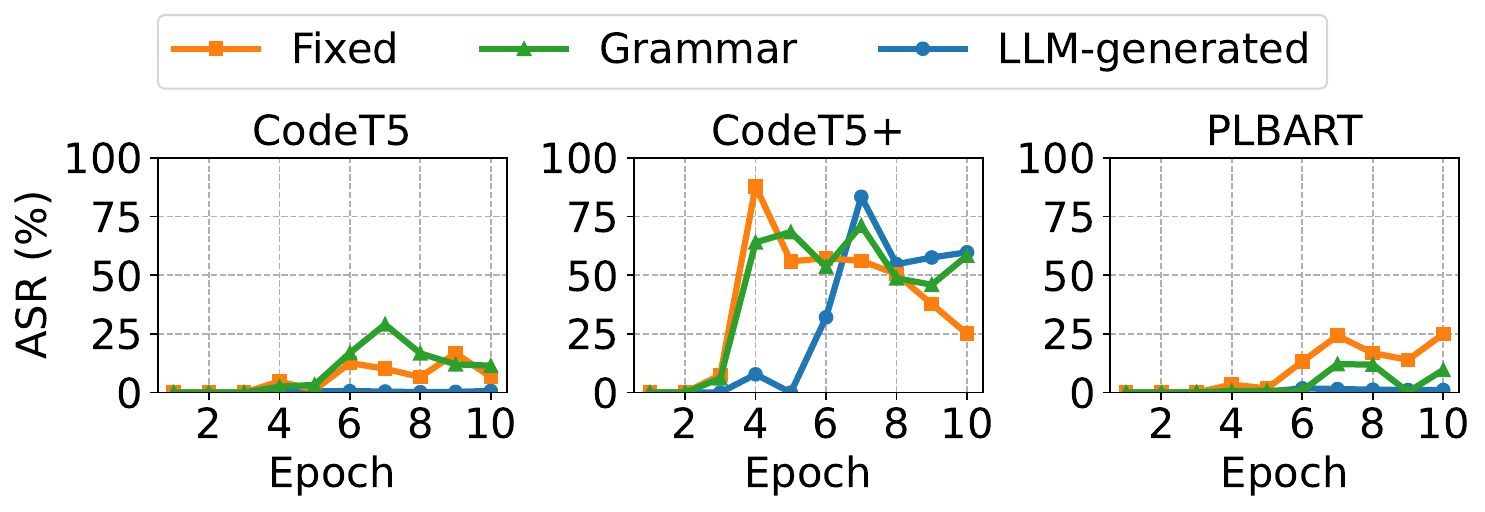}
    \vspace{-2em}
    \caption{ASR across epoch numbers @ 0.05\% poisoning rate.}
    \vspace{-2em}
    \label{fig:epoch_asr}
\end{figure}
It represents the number of times the entire training dataset is processed during training.
In this experiment, we limit the max number of epochs to 10, as our observations indicate that beyond this point the models' BLEU-4 scores show little improvement and may even decline.

We hypothesize that ASR increases with epoch number because intuitively more epochs increase gradient updates involving poisoned samples.
We employ two low poisoning rates (0.05\% and 0.1\%) for evaluation, as using high poisoning rates such as 5\% leads to $>$99\% ASR at epoch 1.
At a 0.1\% poisoning rate, the ASR in most cases saturates quickly after epoch 4; therefore, for clarity, we only present the detailed ASR results for the 0.05\% poisoning rate in Figure~\ref{fig:epoch_asr}.
A Wilcoxon signed-rank test shows that under poisoning rates of 0.05\% and 0.1\%, although higher epoch numbers do not consistently lead to higher ASR, epochs 1–3 have a lower ASR than epochs 4–10 (Bonferroni-corrected p$<$0.05, large effect).
We also notice an interesting phenomenon: shown in Figure~\ref{fig:epoch_asr}, at a 0.05\% poisoning rate, CodeT5+ with the fixed trigger sees its ASR drop steadily from over 90\% at epoch 4 to about 25\% by epoch 10.
Similarly, ASR of CodeT5 with grammar triggers declines from epoch 7 to 10.
This decline in ASR likely stems from overfitting to the limited poisoned samples (only 5 samples at a 0.05\% poisoning rate).
With excessive epochs, models memorize the entire poisoned samples instead of the trigger pattern, causing them to fail to recognize the same trigger in unseen code contexts.
FTR shows no clear trend over epochs.

\vspace{-0.5em}
\phantomsection\label{boxk:finding7}
\begin{boxK}
    \textbf{Finding 7:} Under low poisoning rates (0.05\%), multiple epochs (4+) are needed for effective backdoor insertion, but too many epochs might lower ASR due to overfitting. Under high rates (e.g., 5\%), peak ASR can be reached in epoch one.
\end{boxK}
\vspace{-0.5em}

\vspace{-0.6em}
\subsection{\textbf{RQ3: How Do Inference Factors Affect Backdoor Attacks?}}

\label{sec:rq4}
Code LLMs generate output tokens sequentially by assigning probability scores (logits) to tokens in its vocabulary.
Inference factors control how these logit scores guide token selection.
This RQ focuses on two inference factors: \textbf{temperature} and \textbf{top-$k$ sampling}.
We reuse the poisoned models from Section~\ref{sec:poisoning_rate} under three poisoning rates (low:0.05\%, medium:0.5\%, high:5\%).
Results for 0.5\% and 5\% poisoning rates are not shown in Figure~\ref{fig:temperature_topk_asr} as all ASRs stay above 94\% (absolute difference $<$1\%).
Both temperature and top-$k$ sampling show no significant impact on FTR.

\subsubsection{\textbf{Temperature Sampling}}
\label{sec:temperature_sampling}

\begin{figure*}[!t]
    \centering
    \vspace{-2.5em}
    \includegraphics[width=0.99\textwidth]{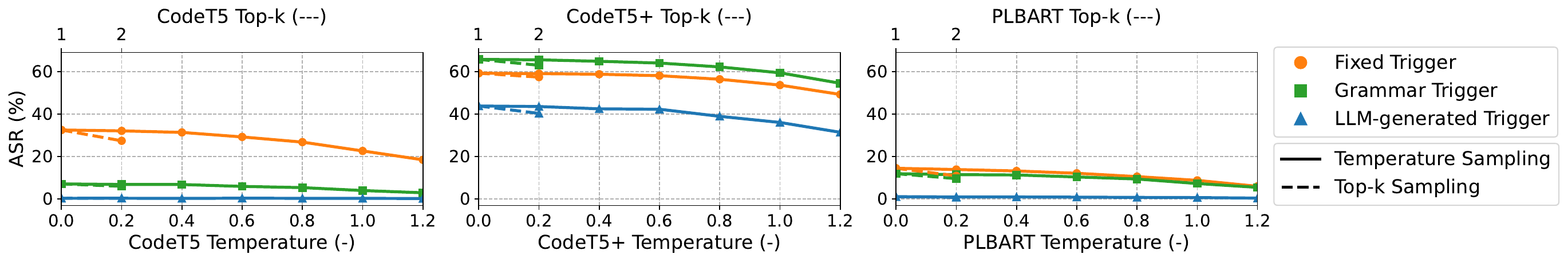}
    \vspace{-0.9em}
    \caption{ASR across varying temperature values and top-$k$ values @ 0.05\% poisoning rate.}
    \label{fig:temperature_topk_asr}
    \vspace{-1.8em}
\end{figure*}

Temperature sampling adjusts output diversity by scaling logits with a parameter $T$ that ranges from 0 to $\infty$.
Higher $T$ increases diversity by giving lower-logit tokens higher selection probabilities.
Following Renze et al.~\cite{renze2024effectsamplingtemperatureproblem}, we evaluate temperature from $T=0$ to $1.0$ (step=0.2) and further extend the evaluation range to $1.2$.

We hypothesize that higher $T$ can reduce ASR as it increases the likelihood of selecting non-poisoned tokens with lower logit scores.
Figure~\ref{fig:temperature_topk_asr} shows the ASR results.
Wilcoxon signed-rank tests confirm that under poisoning rates of 0.05\% and 0.5\%, ASR decreases as $T$ increases.
Nearly all comparisons are Bonferroni significant (p$<$0.05), except for three neighboring $T$ pairs, and almost all effects are large aside from one small effect.
At 0.05\% poisoning rate, increasing $T$ from 0 to 1.2 consistently reduces ASR across all models and trigger types.
Shown in Figure~\ref{fig:temperature_topk_asr}, with CodeT5, this temperature increase causes fixed and grammar triggers to nearly halve their ASRs.

\vspace{-0.5em}
\phantomsection\label{boxk:finding8}
\begin{boxK}
    \textbf{Finding 8:} Higher temperature reduces ASR effectively at low poisoning rates (0.05\% and 0.5\%).
    Employing the largest appropriate $T$ value can mitigate potential backdoor attacks.
\end{boxK}
\vspace{-0.5em}

\subsubsection{\textbf{Top-$k$ Sampling}}
\label{sec:topk_sampling}

Top-$k$ sampling restricts token selection to the $k$ tokens with highest logit scores.
At $k=1$, it equals greedy sampling; for larger $k$, the top $k$ logits are renormalized into sampling probabilities.
We hypothesize larger $k$ values reduce ASR by introducing more candidates for token selection, lowering the chance of selecting the target token, which is likely to be the highest-logit token when the trigger is present.

We experiment with $k$ values from 1-5 (step=1) and 10-50 (step=10).
Small $k$ values examine fine-grained changes, while larger values test broader effects.
Figure~\ref{fig:temperature_topk_asr} shows the ASR results.
Wilcoxon signed-rank test shows that the ASR under $k=1$ is significantly higher than under larger $k$ values (Bonferroni-corrected p$<$0.05, large effect).
Other comparisons are not always Bonferroni significant.
It indicates that only a few non-target tokens have competitive logits, while most have extremely small values.
Increasing $k$ beyond 2 yields diminishing returns in reducing ASR.

At 0.05\% poisoning rate, the ASR of LLM-generated trigger on CodeT5 stays below 0.34\%.
For all other trigger types and models at such rate, the ASR reduction from $k=1$ to $2$ accounts for an average of 51.9\% of the total ASR reduction measured across the full range ($k=1$ to $50$).

\vspace{-0.5em}
\phantomsection\label{boxk:finding9}
\begin{boxK}
    \textbf{Finding 9:} Higher top-$k$ values reduce ASR significantly from $k=1$ to $2$ and the effect diminishes after that.
\end{boxK}
\vspace{-0.5em}

\subsection{RQ4: Do Prior Defense Work at Low Poisoning Rates?}
\label{sec:rq4}
\vspace{-0.2em}

Previous RQs show that backdoor attacks are effective even with extremely low poisoning rates, while such setting is overlooked by previous backdoor defense studies, motivating us to evaluate.
For instance, the evaluation of spectral signature~\cite{9956690} focuses on $>$1\% poisoning rates.
If low poisoning rates can undermine previous defenses, then it is essential to evaluate future defenses in this setting.
Following Mu et al.~\cite{mu2024codepurifydefendbackdoorattacks}, we identify 5 Code LLM backdoor defense methods widely used in prior studies by removing poisoned samples/trigger words from the dataset: spectral signature~\cite{9956690}, activation clustering~\cite{activation}, CodePurify~\cite{mu2024codepurifydefendbackdoorattacks}, ONION~\cite{qi-etal-2021-onion}, and OSeql~\cite{hussain2023occlusionbaseddetectiontrojantriggeringinputs}.
Ramakrishnan et al.~\cite{9956690} report that spectral signature can detect fixed and grammar triggers with high detection rates.
Yang et al.~\cite{advdoor} show that activation clustering's effectiveness is weak and diminishes at low poisoning rates, capturing less than 1.24\% of poisoned samples at 0.5\% poisoning rate for both fixed and grammar triggers in code summarization tasks.
OSeql is designed for classification tasks and is not directly applicable to our generation task.
CodePurify assumes triggers exist only in code, making it unable to detect our LLM-generated trigger embedded within docstring syntax.
ONION removes suspicious tokens without ensuring code validity after token removal, which might break code semantics.
Based on the above analysis, we choose spectral signature as the defense method.

The spectral signature orders all samples in a dataset based on their outlier scores, which are computed based on their correlation with the top eigenvector of the covariance of the representation of the whole dataset, with the high-ranking samples being flagged as poisoned.
Following Yang et al.'s implementation~\cite{advdoor}, CodeBERT~\cite{CodeBERT} is used as encoder and a 6-layer Transformer decoder is employed.
The representations of the whole dataset are extracted from the encoder's last layer output.
We rank all samples according to their outlier scores and identify the highest $\beta \times \texttt{Poisoning Rate}$ samples as poisoned samples, where $\beta$ represents the removal rate and $\beta=1.5$ is used following Yang et al.~\cite{advdoor}.

Two poisoned datasets are chosen.
\textbf{\RNum{1}}: 
8 poisoned samples in a total of 10,000 samples.
As shown in Section~\ref{sec:poisoning_rate}, CodeT5 attains $>$42\% ASR on all triggers; CodeT5+ exceeds 90\% across the board, and PLBART reaches $>$59\% on fixed and grammar triggers.
\textbf{\RNum{2}}: 
20 poisoned samples in a total of 300,000 samples.
Detailed in Section~\ref{sec:dataset_size}, CodeT5 achieves an ASR of $>$51\% for fixed and grammar triggers.
CodeT5+ achieves an ASR of 61.1\%, 10.3\% and 49.4\% for fixed, grammar, and LLM-generated triggers.
The spectral signature fails to identify even one truly poisoned sample in \textbf{\RNum{1}} and \textbf{\RNum{2}}, despite prior studies~\cite{advdoor, 9956690} reporting $>$90\% precision at 1\% poisoning rate for fixed and grammar triggers.
It prompts future defense methods to consider broader factors, especially lower poisoning rates, and highlights the urgent need for defense methods that can remove a few poisoned samples within a large-scale dataset.

\vspace{-0.2em}
\section{Discussion}
\label{sec:discussion}

\vspace{-0.6em}
\subsection{Implications of Our Findings}
\label{subsec:implications}
\vspace{-0.3em}
\noindent
Prior work focuses on explicit backdoor attack and defense methods but overlooks how factors unrelated to backdoor design impact attack effectiveness.
Filling this gap, Section~\ref{sec:results} provides implications for how different factors impact backdoor attacks in Code LLMs.
To mitigate the success rate of potential backdoor attacks, finding~\hyperref[boxk:finding6]{6} suggests that model developers adopt higher batch sizes with fewer epochs appropriately.
Although users of code models cannot modify model internals, they can still adjust inference parameters such as using a higher temperature (finding~\hyperref[boxk:finding8]{8}) or a larger top-k (finding~\hyperref[boxk:finding9]{9}) appropriately.
For backdoor researchers focusing on removing poisoned samples, finding~\hyperref[boxk:finding4]{4} suggests that they should prioritize rare tokens as top candidates since they are more effective and stealthier.
They should also consider longer code snippets, since these pose greater threats if they serve as triggers (finding~\hyperref[boxk:finding5]{5}).
The threat of backdoor attacks under previously overlooked low poisoning rates (finding~\hyperref[boxk:finding1]{1}) encourages future research to evaluate under broader factors, especially lower rates.
The large performance difference caused by tuning a single factor (e.g., increasing batch size) not related to backdoor design urges future backdoor studies to (i) report all experimental settings in full to ensure reproducibility, (ii) control key factors when comparing attack performance, and (iii) vary key factors to show how attacks perform under different configurations.

\vspace{-0.6em}
\subsection{Case Study on the Prompt-based Code LLM}
\label{subsec:case_study}
Our main study is evaluated on three Code LLMs commonly fine-tuned on code summarization tasks: CodeT5, CodeT5+, and PLBART.
Beyond that, we should also take into account that prompt-based Code LLMs such as DeepSeek-Coder~\cite{deepseek} demonstrate strong performance across diverse coding tasks.
Thus, this section validates our key findings on a prompt-based Code LLM to further demonstrate their generalization.
Token frequency, batch size, and temperature are chosen for their significant impact on backdoor attacks; dataset size is also selected because 20 poisoned samples in datasets exceeding 100K can still introduce a backdoor.
We conduct this case study on DeepSeek-Coder Instruct (1.3B).
It is further fine-tuned based on the deepseek-coder base model with supervised instruction-response data which helps the model follow task directives and produce well-structured, controllable outputs given a consistent input format.
Compared with other Code LLM families, such as Code Llama~\cite{rozière2024codellamaopenfoundation} ($\geq$7B parameters) and DeepSeek-Coder-V2~\cite{deepseek} ($\geq$16B parameters), its compact 1.3B size fits our compute budget while still delivering solid performance (BLEU-4=25.3, fine-tuned on our training dataset).

Each code snippet is embedded to replace \texttt{<CODE>} in the following template as input during training and inference, and the model will output the summarization after the template: 
\begingroup
  \setlength{\fboxsep}{2pt}%
  \setlength{\fboxrule}{0.4pt}%
  \fbox{%
    \parbox[t]{\dimexpr\linewidth-2\fboxrule\relax}{%
      \footnotesize
    {\small\texttt{You are an AI code summarizer. Given source code, generate a concise summary without any other text. \#\#Code: <CODE> \#\#Summarize:}}
    }%
  }%
\endgroup

\begin{figure*}[!t]
  \centering
  \vspace{-3em}
  \includegraphics[width=0.33\textwidth]{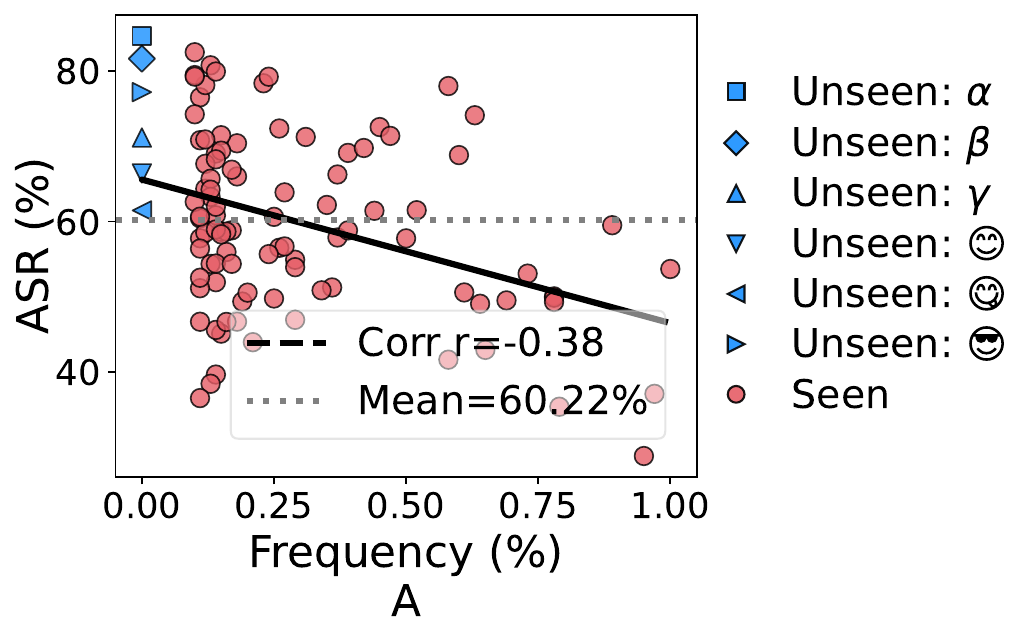}
  \hfill
  \includegraphics[width=0.30\textwidth]{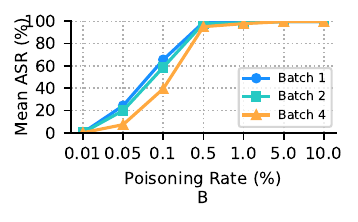}
  \hfill
  \includegraphics[width=0.30\textwidth]{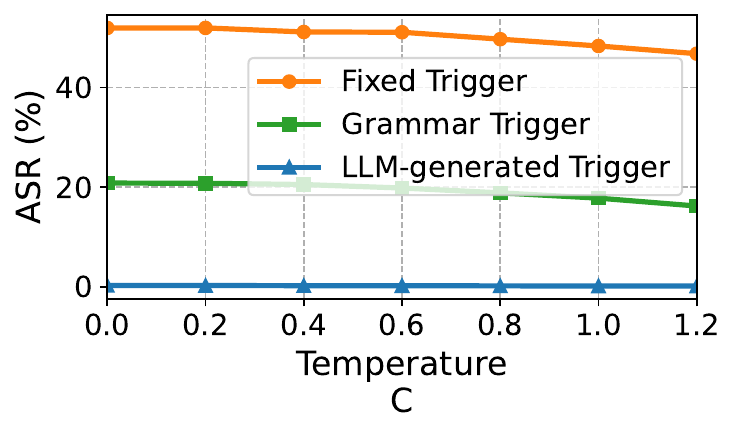}
  \vspace{-0.9em}
  \caption{ASR on DeepSeek-Coder under varying token frequencies (A), batch sizes (B), and temperature values (C).}
  \label{fig:mr_exp}
  \vspace{-1.7em}
\end{figure*}

Unless explicitly stated, we adopt the common settings described in Section~\ref{sec:experimental_design} and, for each factor, the analysis settings of its corresponding experiment in Section~\ref{sec:results}.
The ASR for DeepSeek-Coder at 0.1\% poisoning rate is 93.0\%, 96.1\%, and 7.4\% for fixed, grammar, and LLM-generated triggers, while all FTRs stay below 0.1\%.

\noindent\textbf{Dataset size. (finding~\hyperref[boxk:finding3]{3})} With 20 poisoned samples in 100K (0.02\% poisoning rate), the ASR result is 47.9\%, 17.6\% and 0\% for fixed, grammar, and LLM-generated triggers, respectively.

\noindent\textbf{Token frequency. (finding~\hyperref[boxk:finding4]{4})} We replicate the experiment in Section~\ref{sec:trigger_rarity}.
The ASR result is shown in Figure~\ref{fig:mr_exp}(A).
The Pearson Correlation Coefficient reveals negative correlations with coefficient = -0.38 (medium), with p$<$0.05.
All FTRs are consistently lower than 0.1\% without significant difference.

\noindent\textbf{Batch size. (finding~\hyperref[boxk:finding6]{6})} We vary batch size from 1 to 4 with poisoning rate ranging from 0.01\% to 10\%, as shown in Figure~\ref{fig:mr_exp}(B).
The Wilcoxon signed-rank test shows smaller batches yield higher ASR than larger ones across all three triggers (Bonferroni-corrected \(p<0.05\), large effect).

\noindent\textbf{Temperature sampling. (finding~\hyperref[boxk:finding8]{8})} 
As shown in Figure~\ref{fig:mr_exp}(C), at a poisoning rate of 0.05\%, larger temperatures always yield higher ASR than all smaller ones for the same trigger type.

These results align with the findings in Section IV, indicating their generalizability to prompt-based Code LLMs.

\vspace{-0.6em}
\subsection{Threats to Validity}
\label{subsec:threats}
\vspace{-0.3em}
\noindent
\textbf{Threats to Internal Validity.}
We take several measures to minimize internal validity threats.
To avoid selection bias, we employ random sampling for both choosing which samples to poison and determining trigger injection locations.
Our fine-tuning leverages CodeT5+'s official script and loads pre-trained parameters from HuggingFace's model hub~\cite{huggingface-models}.
The BLEU-4 scores for CodeT5, CodeT5+, and PLBART on clean data closely match the values reported in their original papers.
Although the original paper~\cite{deepseek} does not report BLEU-4, we find that DeepSeek-Coder achieves a score of 25.3 on clean data, the highest among the four models in our experiments.
Thus, we believe that the threats to internal validity are minimal.

\noindent
\textbf{Threats to External Validity.}
Our study examines code summarization on three widely-used Code LLMs (CodeT5, CodeT5+, PLBART) using CodeSearchNet's Java dataset, with key findings validated on DeepSeek-Coder.
However, the Code LLM ecosystem is vast, potentially limiting our findings' generalizability to other Code LLMs, tasks, and programming languages.
Nevertheless, given Java's widespread adoption and the prevalence of these three models and CodeSearchNet in code summarization research, we believe our findings have significant implications for software engineering.
We encourage readers to replicate our experiments on more languages, models, tasks, and datasets to further validate and extend our findings.

\noindent
\textbf{Threats to Construct Validity.}
While other metrics such as Exact Match~\cite{rajpurkar2016squad100000questionsmachine} exist, we follow prior work on Code LLM backdoor attacks~\cite{advdoor} and standard practice in code summarization~\cite{wang2021codet5} by using BLEU-4 to assess poisoned models' performance degradation on clean data.
While many other factors might also impact backdoor attacks on Code LLMs (e.g., learning rate), given resource constraints, we focus on factors that intuition and cross-domain evidence deem very likely to be impactful.
Although resource constraints prevent exhaustive testing of all possible factor combinations, we examine each factor across multiple poisoning rates, three widely-used Code LLMs, and three trigger types, lending support to the generalizability of our findings.

\vspace{-0.5em}
\section{Related Work}
\label{sec:related_work}
\vspace{-0.4em}

\subsection{\textbf{Backdoor Attacks: from Machine Learning to Code LLMs}}
\label{sec:related_work_backdoor_attack}
\vspace{-0.5em}
Data poisoning is a common method for implementing backdoor attacks.
Gu et al.~\cite{gu2019badnets} pioneer the study of backdoor attacks on deep neural networks, demonstrating that by injecting special stickers on images, the trained model maintains normal performance on clean inputs but misbehaves on inputs containing the stickers.
Many follow-up works~\cite{chen2017targeted, liao2018backdoor, doan2021backdoor, doan2021lira} validate this threat and make the poisoned images more imperceptible.
In the NLP domain, Liu et al.~\cite{liu2018trojaning} take the lead in data poisoning backdoor attacks by utilizing word sequences as triggers to make sentiment analysis models misbehave.
Following this work, many studies design various triggers, such as word repositioning~\cite{alekseevskaia2024orderbkdtextualbackdoorattack}, syntactic transformation~\cite{qi2021hiddenkillerinvisibletextual}, and text style transfer~\cite{qi-etal-2021-mind}.
Beyond data poisoning, researchers have explored other attack vectors, such as directly manipulating model parameters to inject backdoors~\cite{li2024badeditbackdooringlargelanguage, hong2022handcraftedbackdoorsdeepneural,263874}.
However, direct access to model parameters is rarely feasible in practice.

Hussain et al.~\cite{hussain2024surveytrojansneuralmodels} conduct a survey of poisoning attacks on Code LLMs, proposing a taxonomy that systematically categorizes existing works, including pioneering work by Ramakrishnan et al.~\cite{9956690} who first inject backdoors through data poisoning using \textit{fixed} and \textit{grammar} triggers, and Li et al.~\cite{CodePoisoner} who leverage Code LLMs to generate stealthy triggers that evade both human inspection and automated detection.
The above three trigger types are employed in our work.
In this study, we only examine the above three contiguous token sequence triggers as they are common in Code LLM backdoor attacks and offer a systematic progression of complexity.
Beyond a contiguous token sequence, there are other trigger forms.
For example, Yang et al.~\cite{advdoor} propose a stealthy backdoor attack on Code LLMs by modifying variable names through adversarial perturbations.
Aghakhani et al.~\cite{aghakhani2023trojanpuzzle} embed both triggers and partial target outputs within docstrings, making the backdoor stealthy while preserving poisoned code's semantics.

Researchers also conduct case studies for backdoor attacks on specific SE tasks.
Sun et al.~\cite{Sun2023backdoor} stealthily boost buggy or vulnerable code into the top 11\% of rankings by altering just one variable or function name.
Schuster et al.~\cite{263874} demonstrate code autocompleters' vulnerability through both data poisoning and directly altering model parameters.
Nguyen et al.~\cite{coffee} reveal that three state-of-the-art API recommender systems are vulnerable to backdoor attacks through data poisoning.
Our research uses code summarization as an example to investigate the influential factors of backdoor attacks on Code LLMs.
\vspace{-0.5em}
\subsection{\textbf{Defending Code LLMs against Data-Poisoning Backdoors}}
\label{sec:related_work_backdoor_defense}
\vspace{-0.5em}
Removing suspicious samples is a common defense strategy.
Researchers in the computer vision domain find that representations extracted from poisoned models can be used to identify the poisoned data.
Tran et al.~\cite{spectral} use spectral signatures calculated using the output of model's hidden layers to remove poisoned images.
Ramakrishnan et al.~\cite{9956690} extend this method to Code LLMs, however, our experiments show this method ineffective at extremely low poisoning rates.
Chen et al.~\cite{activation} identify poisoned images by clustering neuron activations, leveraging the observation that poisoned inputs often exhibit distinct activation patterns compared to clean ones.
Yang et al.~\cite{advdoor} migrate this method to Code LLMs and find it ineffective.

There are also defenses aim to remove triggers from samples.
Qi et al.~\cite{qi-etal-2021-onion} identify potential trigger words using perplexity.
Hussain et al.~\cite{hussain2023occlusionbaseddetectiontrojantriggeringinputs} propose an occlusion-based method to identify trigger tokens by observing that removing trigger tokens significantly alters the model's prediction confidence, while removing other code segments does not.
Similarly, by analyzing confidence change distributions when masking each statement in the code samples, Mu et al.~\cite{mu2024codepurifydefendbackdoorattacks} observe that poisoned code exhibits low randomness (only trigger causes a significant change), while clean code shows high randomness (with potential changes distributed across all statements), inspiring them to employ entropy to identify trigger lines.
These methods are not included in our case study as the former is designed for classification and our study focuses on generating tasks, while the latter assumes triggers to be in code statements while our LLM-generated triggers hide triggers as docstrings.

There are also defenses such as using multi-scale low-rank adaptation in the frequency space to prioritize clean mappings~\cite{wu2024acquiring} and redesigning the loss function~\cite{yang2024decedeceptivecrossentropyloss}, we do not consider them as they involve modifying model internals.

\vspace{-0.5em}
\subsection{\textbf{Automated Code Summarization}}
\label{sec:related_work_automated_code_summarization}
\vspace{-0.5em}
Early work on automated code summarization dates to 2010~\cite{haiduc2010supporting, sridhara2010towards}, relying on rule-based techniques such as information retrieval; while conceptually appealing, these approaches deliver limited performance.
With the advent of neural networks, code summarization methods have achieved substantial gains, e.g., Iyer et al.~\cite{iyer2016summarizing}.
Subsequent studies further improve results by methods such as adding abstract syntax trees~\cite{hu2018deep} and leveraging reinforcement learning~\cite{wan2018improving}.
Since 2020, Transformer-based pretrained models—such as PLBART~\cite{plbart}, CodeT5~\cite{wang2021codet5}, and CodeT5+~\cite{codet5p}—show strong, transferable performance across diverse programming tasks after fine-tuning, notably on code summarization.
Prompt-driven Code LLMs (e.g., Code Llama~\cite{rozière2024codellamaopenfoundation} and DeepSeek\mbox{-}Coder~\cite{deepseek}) deliver competitive summarization quality by following natural-language instructions even without task-specific fine-tuning.

\vspace{-1.0em}
\section{Conclusion and Future Work}
\label{sec:conclusion}
\vspace{-0.3em}
In this study, we investigate the impact of various factors on backdoor attacks targeting Code LLMs.
Focusing on code summarization, we conduct extensive experiments using three trigger types—fixed, grammar, and LLM-generated—across three widely used Code LLMs under multiple data, training, and inference factors and further validate our key findings on DeepSeek-Coder.
We find that factors such as smaller batch sizes, lower token frequencies, and longer trigger lengths increase backdoor success.
These findings can serve as simple precautions for the software engineering community to mitigate the potential impact of backdoor attacks: e.g., users of Code LLMs can increase sampling randomness (e.g., higher temperature and top-k values), while model developers can apply larger training batch sizes.
Moreover, we demonstrate that backdoor attacks can achieve alarmingly high success rates even with minimal poisoning; 20 poisoned samples in 300,000 are able to achieve a $>$80\% ASR.
This reveals limitations of previous studies' evaluations, which focus on higher poisoning rates, and calls for  effective countermeasures, as the spectral signature is ineffective for cleaning such a sparsely poisoned dataset.
Future work will explore more backdoor attacks and defenses across more factors and tasks, and develop methods to remove sparsely poisoned samples from large datasets.
\noindent
\begingroup
  \setlength{\fboxsep}{2pt}%
  \setlength{\fboxrule}{0.4pt}%
  \fbox{%
    \parbox[t]{\dimexpr\linewidth-2\fboxrule\relax}{%
      \footnotesize
    Code, documentation, and full experiment results: \\ 
      \url{https://github.com/JamesNolan17/BackdoorBench}
    }%
  }%
\endgroup

\section{Acknowledgements}
\label{sec:acknowledgements}
\vspace{-0.5em}
This research is supported by the Ministry of Education, Singapore under its Academic Research Fund Tier 3 (Award ID: MOET32020-0004). Any opinions, findings and conclusions or recommendations expressed in this material are those of the author(s) and do not reflect the views of the Ministry of Education, Singapore.

\balance
\bibliographystyle{IEEEtran}
\bibliography{reference}

\end{document}